\documentstyle[12pt]{article}
\title{Phase diagrams of a Spin-1 Ising Superlattice }
\date{April 29, 2004}
\author{Hamid EZ-ZAHRAOUY$^{(a)} $\footnote{Corresponding author: ezahamid@fsr.ac.ma} and Ahmed KASSOU-OU-ALI$^{(b)}$}

\begin{document}
\maketitle
\begin{center}
{\it \small
(a) Facult\'e des Sciences, D\'epartement de Physique, Laboratoire de Magn\'etisme et Physique des Hautes Energies, B.P. 1014, Rabat, Morocco.\\
(b) Facult\'e des Sciences, D\'epartement de Physique, Laboratoire de Physique Th\'eorique, B.P. 1014, Rabat, Morocco.}
\end{center}

\abstract{The three-dimensional spin-1 Ising superlattice consisting of two different ferromagnetic materials with two different crystal fields $\Delta_1$ and $\Delta_2$ is considered in the mean field approximation. The phase diagrams are considered in the (t,$d_2$) plane for different ranges of variation of $d_1$ (t=T/J, $d_1$/J and $d_2$/J are the reduced temperature and crystal fields respectively). The phase diagrams exhibit a variety of multicritical points and reentrant and double reentrant behaviours. They are found to depend qualitatively and/or quantitatively on the thicknesses of the materials in a supercell. This has direct consequences on the nature of the magnetic states of superlattices with different thicknesses.
\newpage
\section{Introduction}
\setlength{\parskip}{.2in} 
The magnetic properties of magnetic thin films and multilayers have attracted increased attention in recent years [1]. This is because of the easiness of their preparation made possible by the recent advances of modern vacuum science, in particular the molecular beam epitaxy technique, the novel magnetic phenomena they exhibit compared to bulk materials, and their potential technological importance. The unusual magnetic properties of these artificial materials are mostly attributed to surface and interface effects, to the reduced dimensionality and to magnetic interactions.

The thickness dependence of most of the properties of these materials is now an established fact. For instance, magnetization measurements of a series of Fe films [2] show that the magnetic moment of Fe depends on the thickness of the film. In ultrathin Ni films grown on Cu(001) [3], a sharp spin reorientation transition at small film thicknesses (from 5 to 7 monolayers) from in-plane to perpendicular magnetization was observed. At much higher thicknesses (from 35 to 70 monolayers) the magnetic moments reorient gradually and become again parallel to the surface. In superlattices with magnetic layers separated by non magnetic spacer layers, the coupling between two magnetic layers oscillate generally in sign as a function of the thickness of the spacer. This is the case, for example, of the epitaxially grown Gd/Y superlattices [4]. Also, a systematic study of EuTe/PbTe superlattices [5], shows considerable changes of the Néel temperature depending on the number of monolayers of EuTe and PbTe in a supercell.

 The experimental activity on these new materials has been accompanied by significant theoretical effort in order to understand their novel properties. Besides, theoretical work has been devoted to the investigation, on thin film and multiplayer geometries, of known interesting models. This has led to the prediction of new interesting physics. 
 
The Blume-Capel (BC) model for spin 1 is an Ising system which takes into account the single-ion crystal field anisotropy. It has been investigated in detail using many approximate methods, namely mean field approximation [6,7], high temperature series expansion [8], constant-coupling approximation [9], Monte-Carlo [10] and renormalization-group [11] techniques. All of these approximate schemes suggest the existence of a tricritical point in the phase diagram of the model and the instability of the ordered phase beyond a critical value of the anisotropy parameter. On the other hand the BC model has been investigated on semi-infinite lattices with modified surface coupling. Several approximate methods have been used [12] (and references therein). All of  them show the possibility to have a phase with ordered surface and disordered bulk, and show the existence of the so-called 'extraordinary', 'surface' and 'ordinary' transitions, and of multicritical points called special points [13]. More recently, with the increased interest in thin films and multilayers, many works have been devoted to the influence of the crystal field on the phase transitions of magnetic films [14-15] and superlattices [ [16-17].
 
In an earlier paper [18], we considered an Ising spin-1 multilayer (having L layers) with an alternating crystal field: $\Delta_1$ on the odd number layers and $\Delta_2$ on the even number ones. A rich phase transition behavior has been reported there and some novel properties( such as the absence of the tricritical point in the paramagnetic-ferromagnetic transition line and the stability of the ferromagnetic phase at low temperatures for some regions of the parameters of the model, the appearance of the reentrant and double reentrant phenomena and the presence of multicritical and critical end-points) have been found. In the present work, we address our attention to the physical question of the thickness dependence of all those results. The system we consider is a superlattice consisting of two different ferromagnetic materials A and B of different crystal fields: $\Delta_1$ in A and $\Delta_2$ in B. The supercell contains two slabs of materials A and B having different thicknesses: M and N layers respectively. To get rid of  complications due to effects which we don't want to take into consideration (boundary conditions at the surface layers and the parity of the total number of the slabs of the two materials) we consider a three-dimensional superlattice. We find that many of the results mentioned above change qualitatively and /or quantitatively with the variation of the thicknesses of the slabs in a supercell, and that novel phenomena may take place. The rest of the paper is organized as follows. In section 2 we present the model and give the mean field equations of the different order parameters. In section 3 we present the results. Section 4 is devoted to a conclusion.

\section{Model and method}
The system we consider (Fig.1) is a superlattice consisting of two different ferromagnetic materials A and B stacked alternately: the unit cell consists of an arbitrary number M of layers of matarial A with the crystal field $\Delta_1$, and an arbitrary number N of layers of material B with crystal field $\Delta_2$. For simplicity, we focus our attention on the simple cubic structure. The Hamiltonian of this system is given by
\begin{equation}                                 
{\cal H}=-J\sum_{\langle i,j \rangle}S_iS_j+\sum_i \Delta_i S_i^2  
\end{equation}                                   
where $S_i=\pm1,0$. The first sum runs over all pairs of nearest-neighbors and $\Delta_i$ is the crystal field which takes the value $\Delta_1$ on the layers of A and $\Delta_2$ on the layers of B. The periodic condition suggests that we have to consider only the unit cell of thickness L=M+N. The mean field equations of state are straightforwardly obtained and are given by 
\begin{equation}                                 
m_k=\langle S_k \rangle={2\sinh \beta h_k \over e^{\beta \Delta_k}+2 \cosh \beta h_k}
\label{mk}
\end{equation}                                   
\begin{equation}                                 
q_k=\langle S_k^2 \rangle={2\cosh \beta h_k \over e^{\beta \Delta_k}+2 \cosh \beta h_k}
\end{equation}                                   
with 
\begin{equation}                                      
h_k=J(m_{k-1}+zm_k+m_{k+1}) \makebox{, for } k=1,...,M+N\\     
\end{equation}                                 
and  
\begin{equation}                                 
h_{k+M+N}=h_k   
\end{equation}
to ensure the periodic condition 
\begin{equation}                                    
m_{k+M+N}=m_k
\end{equation}   
In these equations, $m_k$ ($q_k$) is the reduced magnetization (quadrupolar moment) of the $k^{th}$  layer. z is the interlayer coordination number (z=4 for the case of the square layers considered here). The reduced total magnetization is $m={1 \over N}\sum_{k=1}^N m_k$. The reduced free energy of the film is given by
\begin{eqnarray}
f=-{1\over \beta} \sum_{k=1}^{M+N} \ln \left[ 1+2e^{-\beta \Delta_k} \cosh \beta h_k \right]+{1 \over 2}\sum_{k=2}^{M+N}(m_{k-1}+zm_k+m_{k+1})m_k \nonumber \\
\label{fren}
\end{eqnarray}                                 

\section{Results and discussions}

We will be interested in the phase diagrams in the (t, $d_2$) plane for different ranges of variation of $d_1$ ($d_1=\Delta_1/J$, $d_2=\Delta_2/J$ and $t=k_{B}T/J$ are the values of the reduced crystal fields and temperature respectively). In order to determine exhaustively the different phase diagrams, we first determine exactly the ground state phase diagram in the  ($d_1,d_2$) plane by looking for the lowest energy configurations. For arbitrary numbers of layers M and N of the slabs of the two materials A and B in the unit cell, we distinguish four regions of this plane (Fig.2):
\begin{itemize}
\item[(a)] $d_1<3-1/M$, $d_2<3-1/N$ and $M(d_1-3)+N(d_2-3)<0$. The stable configuration is $1^M1^N$, where the spins of the sites of the M layers of the slab A and the N layers of the slab B are all equal to 1.
\item[(b)] $d_1<3-1/M$, $d_2>3-1/N$. The stable configuration is $1^M0^N$, where the spins of the slab A are all equal to 1 while those of the slab B are all equal to 0.
\item[(c)] $d_1>3-1/M$ and $d_2<3-1/N$. The stable configuration is $0^M1^N$ .
\item[(d)] $d_1>3-1/M$, $d_2>3-1/N$ and $M(d_1-3)+N(d_2-3)>0$. The stable configuration is $0^M0^N$.	
\end{itemize}

One of the specificities of the superlattice geometry is worth to note here. Material A in bulk with $d_1>3$ is non magnetic at zero temperature, but inside a superlattice it may be ferromagnetic up to $d_1=3+1/M$ (Fig.2). This is an interaction and reduced dimensionality effect: $d_1$ may take higher and higher values for thinner and thinner slabs of material A. 

From Fig.2 it is clear that for fixed values of $d_1$, the system exhibits various transitions by varying $d_2$. For the following ranges of variation of $d_1$: i) $0<d_1<3-1/M$, ii) $3-1/M<d_1<3+1/M$, iii)$d_1>3+1/M$, the system presents the following phase transitions respectively:
\begin{itemize}
\item[i)] from the state $1^M1^N$ to the state $1^M0^N$ at $d_2=3+1/N$.
\item[ii)] from the state $1^M1^N$ to the state $0^M0^N$ at $N(d_2-3)=-M(d_1-3)$.
\item[iii)] from the state $0^M1^N$ to the state $0^M0^N$ at $d_2=3-1/N$.     
\end{itemize}

The topology of the ground state phase diagram is the same for arbitrary values of M and N (Fig.2). Nevertheless, the regions defining the different ground states in the $(d_1,d_2)$ plane are M and N dependant. This implies possible phase transitions, at zero temperature, due to thickness variations. For example, for $d_1=2.6$ and $d_2=2.85$, the system with $(M=1,N=2)$ is non magnetic. With increasing the thickness of the layers of material A, the system ($(M=2,N=2)$ for example) becomes ferromagnetic.This is formally due to the variation of the slope of the line (oblic segment of equation $M(d_1-3)+N(d_2-3)=0$) separating the ferromagnetic from the the non magnetic states, when M or N change.  

For finite temperatures the transition lines are determined by solving numerically Eq.(2) which may have more than one solution. The one which minimizes the free energy (Eq.(7)) corresponds to the stable phase. Transitions of the second order are characterized by a continuous vanishing of the magnetizations while those of the first order exhibit discontinuities at the transition points.

In order to describe the different entities in the phase diagrams, the Griffiths notations [19] will be adopted: the critical end-point $B^mA^n$ denotes the intersection of $m$ lines of second order and $n$ of first order. The multicritical point $B^m$ denotes the intersection of m lines of second order. The tricritical point, which is the intersection of a line of second order and a line of first order, is denoted in particular by $C$. 

In addition, each phase will be represented by the juxtaposition of two expressions each one giving the state of a slab A or B in the unit cell. The notation $(F^M)(F^p0^{N-2p}F^p)$ means that the slab A is in the state $(F^M)$ (i.e. its M layers are all in ferromagnetic states), and the slab B is in the state $(F^p0^{N-2p}F^p)$ (i.e. the first p layers of B, and the last p ones because of symmetry, are all in ferromagnetic states while the $N-2p$ inner ones are all in the non magnetic state 0). A layer in the non-magnetic state 0 has a vanishing magnetization and quadrupolar moment. In the paramagnetic state P only the quadrupolar moment is non vanishing, and in the ferromagnetic state F both quantities are non vanishing.  

In what follows, we will give the phase diagrams in the (t,$d_2$) plane for given values of M and N ($M=5$, $N=9$). Meanwhile we give the phase diagrams for other values of M or N if their topologies are different from those of ($M=5$, $N=9$). For arbitrary fixed values of M and N, the different phase diagrams may be gathered into four groups, each one is characterized by a certain range of variation of $d_1$:

1) For $0<d_1<d_{trc}^{(M)}$ (e.g. Fig.3(a)), the system presents the paramagnetic phase $(P^M)(P^N)$ and the ferromagnetic ones $(F^M)(F^N)$ (called the ordered phase) and $(F^M)(F^p0^{N-2p}F^p)$ (called the partially ordered phases). The system transits from one  ferromagnetic phase to another when a certain layer, and its symmetric, transits from the non magnetic state 0 to a ferromagnetic one F. There is a critical line separating the paramagnetic phase $(P^M)(P^N)$ from the ferromagnetic ones. The ordered phase is separated from the partially ordered ones by a transition line of first order at low temperatures, which meets a critical line at a tricritical point. The critical line meets the paramagnetic-ferromagnetic (P-F) transition line at a multicritical point of the type $B^3$. The partially ordered phases are separated by critical lines, each one representing the transition of a layer, and its symmetric, from a non magnetic to a magnetic state. These lines end, at low temperatures, at some tricritical points and end at some multicritical points of the type $B^3$ on the P-F transition line. They furthermore present the reentrant behavior near the P-F transition. Note that the first order line, which meets the $d_2$ axis at $d_2$=3+1/N, presents an inclination inside the ordered phase $(F^M)(F^N)$. This means that, contrary to the zero temperature case, some layers of the  slab B may be non magnetic even for $d_2$ less, but close to 3+1/N, at low temperatures. This also means the existence of the reentrant behavior where for $d_2$ close to 3+1/N the system, at the beginning in the ordered phase $(F^M)(F^N)$, transits to some partially ordered phases and then returns to the ordered one. This is clarified by the inset in Fig.3(a). It has been found that this behavior does not exist for small values of N (e.g. N=2) 
in which case the first order line mentioned above is parallel to the t-axis. The topology described here is characterized, contrary to the ordinary Blume-Capel model, by the absence of a tricritical point in the (P-F) transition line. The value $d_{trc}^{(M)}$ of $d_1$ above which this point appears is found to be the mean field value of the tricritical crystal field of the spin-1 Blume-Capel film with M layers and a uniform crystal field. $d_{trc}^{(M)}$ increases from $d_{trc}^{(1)}=2.$ for one layer to $d_{trc}^{(\infty)}=3.$ for the bulk. The M dependance of $d_{trc}^{(M)}$ implies that for particular values of $d_1$, superlattices with M and M' so that $d_{trc}^{(M)}<d_1<d_{trc}^{(M')}$, have qualitatively different phase diagrams. For example, for $d_1=2.62$, the system with thickness $M=4$ ($d_{trc}^{(4)}=2.59$) presents a tricritical point, but not the system with $M=5$ ($d_{trc}^{(5)}=2.65$). The absence of the tricritical point in the phase diagrams of the Ising mixed $spin-1/2,spin-1$ superlattice of [17], may be considered as the particular case $d_1=0.$ of the present model.

2) For $d_{trc}^{(M)}<d_1<3-1/M$, three types of topology are found with increasing $d_1$, all of them present a tricritical point in the P-F transition line and present the same features of the type-1 topology with some differences. For $d_1$ in the vicinity of $d_{trc}^{(M)}$ some of the lines of transition (F-F lines) separating the ferromagnetic phases end at some multicritical points of the type $B^3$ on the (P-F) transition line, but some others end at points of the type $BA^2$ [e.g. Fig.3(b)]. For higher values of $d_1$, all the F-F lines end at the multicritical lines of the type $BA^2$ on the P-F line [e.g. Fig.3(c)]. Finally for $d_1$ close to (3-1/M), some partially ordered phases disappear ($(F^5)(F^40F^4)$ in the case of M=5, N=9). This is true only for thick slabs of material B and means that the inner layers of this material can not be driven to the ferromagnetic state by the temperature. In the other hand, the transition line separating the ordered and the partially ordered phases is now of totally first order, and meets the (P-F) line at a triple-point $A^3$ [e.g. Fig.3(d)]. Such topology does not exist for small values of N ($N\leq7$).

For $d_1<3-1/M$ the ferromagnetic phase is stable even for an infinitely large value of $d_2$ at temperatures lower than a temperature ${t_c}^{(M)}(d_1)$ which is independent of the thickness N of the slab B. ${t_c}^{(M)}(d_1)$ is but the mean-field critical temperature of the spin-1 Blume-Capel film with M layers and a uniform crystal field that is equal to $d_1$. This may be understood beyond the mean-field approximation [13]. ${t_c}^{(M)}(d_1)$ is a decreasing function of $d_1$. This means that for large values of $d_2$ the (P-F) transition line takes place at lower temperatures for higher values of $d_1$. In fact, as is clear from Figs.2(a-d), this is true for all $d_2$. This fact explains the progressively disappearance of the re-entrance in the lines separating the ferromagnetic phases near the (P-F) transition line, the eventual disappearance of some partially ordered phases, and the eventual appearance of the $A^3$ point.

For $d_1>3-1/M$, the ferromagnetic phase is unstable for large values of $d_2$, as will be seen below, in accordance with the ground-state phase diagram. This implies that the magnetic state at low temperatures is M dependant. For example, for $d_1=2.75$, the system with $(M=5,N=4)$ exhibits a ferromagnetic order at low temperatures for arbitrary values of $d_2$, but not the system with $(M=2,N=4)$.

3) For $3-1/M<d_1<3+1/M$, no layers transitions may take place. The phase diagram for $(M=5,N=9)$ is of the same type as that of the ordinary spin-1 BC model [e.g. Fig.3(e)]. The first order line reaches the $d_2$ axis at $d_2$ given by $N(d_2-3)=-M(d_1-3)$. Nevertheless, a reentrants and double reentrants behaviors may appear in the P-F transition line for varied values of the thicknesses. This is the case for values of $d_1$ close to $(3-1/M)$ for small values of M relatively to N (e.g. Fig.4(a) for M=2 and N=9), and for values of $d_1$ close to $(3+1/M)$ for high values of M relatively to N (e.g. Fig.4(b) for M=9 and N=2).  

4) For $d_1>3+1/M$ many types of topology are found for M=5 and N=9, all of them have a (P-F) transition line containing a tricritical point C and meet the $d_2$ axis at $d_2=3-1/N$. Some of the ferromagnetic phases $(0^5)(F^9)$, $(F0^3F)(F^9)$, $(F^20F^2)(F^9)$ or $(0^5)(F^9)$ may appear in the phase diagrams depending on the values of $d_1$. For $d_1$ close to (3+1/M), all of these phases exist and are separated by critical lines of approximately constant temperatures which meet the P-F transition line at three critical end points of the type $(BA^2)$ [e.g. Fig.3(f)]. With increasing $d_1$, the following changes take place,

-the ferromagnetic phase $(F^5)(F^9)$ becomes totally inside the phase $(F^20F^2)(F^9)$ which exhibits an island behavior [e.g. Fig.3(g)].

- The phase $(F^20F^2)(F^9)$ becomes itself totally inside the phase $(F0^3F)(F^9)$ and two islands then appear [e.g. Fig.3(h)].

-The ferromagnetic phase $(F^5)(F^9)$ disappears from the phase diagram [e.g. Fig.3(i)]

- The phase $(F^20F^2)(F^9)$ disappears too, and only the two phases $(0^5)(F^9)$ and $(F0^3F)(F^9)$ still exist. They are separated by a critical line which meets the (P-F) transition line at a ($BA^2$) point [e.g. Fig.3(j)].

- The latter line exhibits a reentrant behavior and the phase $(F0^3F)(F^9)$ becomes totally inside the $(0^5)(F^9)$ phase [e.g. Fig.3(k)].

- The phase $(F0^3F)(F^9)$ disappears and the topology becomes of the ordinary BC type, but now with the partially ordered phase $(0^5)(F^9)$ at low temperatures, instead of the ordered phase $(F^5)(F^9)$ [e.g. Fig.3(l)].

For general values of M and N, analogous phase diagrams are found, the number of the ferromagnetic phases which appear in these diagrams depends obviously on M and N and depends on $d_1$. For some values of M and N, the (P-F) transition line may exhibit the reentrant or double-reentrant behaviors for $d_1$ close to 3+1/M. This is the case, for example, for M=9 and N=2 (Fig.4(b)).
	 
\section{Conclusion}

We have studied, using the mean field theory, the phase diagrams of a spin-1 Blume-Capel superlattice consisting of two different ferromagnetic materials with two different crystal fields $\Delta_1$ and $\Delta_2$. The ground state phase diagram in the ($d_1,d_2$) plane was determined exactly for arbitrary values of the thicknesses M and N of the constituents in a unit cell. At finite temperature, the phase diagrams have been given in the (t,$d_2$) plane for different ranges of variation of $d_1$ .They exhibit a variety of multicritical points, in particular the usual tricritical point C in the paramagnetic-ferromagnetic line of transition. This point does not appear for values of $d_1$ below a threshold value $d_{trc}^{(M)}$, which depends only on the thickness M of the constituent with crystal field $\Delta_1$ in the unit cell. Moreover, lines of transition presenting the reentrant, double reentrant and island behaviors may also take place. The phase diagrams depend qualitatively and/or quantitatively on M and N. This has direct consequences on the magnetic nature of superlattices with different thicknesses.

\newpage
\section*{References}

\begin{description}
\item[][1] For a review see R.E. Camlay and R. L. Stamps, , J. Phys.: Condens. Matter {\bf 5}, 3727 (1996)
\item[][2] Yi Li, C. Polaczyk, F. Klose, J. Kapoor, H. Maletta, F. Mezei, and D. Riegel, Phys. Rev. B {\bf 53}, 5541 (1996)
\item[][3] B. Schulz and K. Baberschke, Phys. Rev. B {\bf 50}, 13467 (1994)\\
B. Schulz, R. Schwarzwald and K. Baberschke, Surf. Sci. {\bf 307-309}, 1102 (1994) \\
W. L. O'Brien and B. P. Tonner, J. Appl. Phys. {\bf 79}, 5623 (1996)\\
W. L. O'Brien, T. Droubay and B. P. Tonner, J. Appl. Phys. Rev. B {\bf 54}, 9297 (1996)
\item[][4] S. S. P. Parkin, N. More , K. P.Roche, Phys.Rev.Lett. {\bf 64}, 2304 (1990)
\item[][5] H. Kepa, G. Springholz, T. M. Giebultowicz, K. I. Goldman, C. F. Majkrzak, P. Kacman, J. Blinowski, S. Holl, H.            Krenn, and G. Bauer Phys. Rev. B {\bf 68}, 024419 (2003)
\item[][1-6] M. Blume, Phys. Rev. {\bf 141}, 517 (1966)
\item[][2-7] H.W. Capel, Physica {\bf 32}, 966 (1966)
\item[][8] D.M. Saul and M. Wortis, Amer. Inst. Phys. Conf. Proc. {\bf 5}, 349 (1972)\\
    P.F. Fox and D.S. Gaunt, J. Phys. C {\bf 5}, 3085 (1972)\\
    D.M. Saul, M. Wortis and D. Stauffer, Phys. Rev. B {\bf 9}, 4964 (1974)
\item[][9] M. Tanaka and K. Takahashi, Phys. Stat. Sol. (b) {\bf 93}, K85 (1979)
\item[][10] B.L. Arora and D.P. Landau, Amer. Inst. Phys. Conf. Proc. {\bf 5}, 352 (1972)\\
    A.K. Jain and D.P. Landau, Phys. Rev. B {\bf 22}, 445 (1980)\\
    W. Selke and J. Yeomans, J. Phys. A {\bf 16}, 2789 (1983)\\
    C.M. Care, J. Phys. A {\bf 26}, 1481 (1993)\\
    M. Deserno, Phys. Rev. E {\bf 56}, 5204 (1997)
\item[][11] A.N. Berker and M. Wortis, Phys. Rev. B {\bf 14}, 4946 (1676)\\
      T.W. Bukhardt, Phys. Rev. B {\bf 14}, 1196 (1976)\\
      H. Dickinson and J. Yeomans, J. Phys. C16, L345 (1983)
\item[][12] C. Buzano and A. Pelizzola, Physica A {\bf 216}, 158 (1995)
\item[][13] A. Benyoussef, N. Boccara and M. Saber, J. Phys. C {\bf 19}, 1983 (1986)
\item[][14] T. Balcerzak, J. Magn. Magn. Mat. {\bf 137}, 57 (1994)
\item[][15] L. Bahmad, A. Benyoussef and H. Ez-Zahraouy, J. Magn. Magn. Mat. {\bf 251}, 115 (2002)
\item[][16] G.Wiatrowski, G.Bayreuther and F.Bensch, Physica A {\bf 293}, 478 (2001)
\item[][17] E. F. Sarmento, J. C. Cressoni, and R. J. V. dos Santos, J. Appl. Phys. {\bf 75}(10), 5820 (1994)
\item[][18] H. Ez-Zahraouy and A. Kassou-Ou-Ali, Phys. Rev. B {\bf 69}, 64415 (2004)
\item[][19] R.B. Griffiths, Phys. Rev. B {\bf 12}, 345 (1975)

\end{description}

\newpage
\section*{Figure captions}

\begin{description}
\item[Fig.1] A schematic illustration of the model.
\item[Fig.2] The ground state phase diagram in the $(d_1,d_2)$ plane for arbitrary numbers M and N of the layers of the slabs A and B respectively, in a unit cell. 
\item[Fig.3] The phase diagrams in the (t,$d_2$) plane for $M=5$ and $N=9$ and generic values of $d_1$. (a) $d_1=1$., (b) $d_1=2.66$, (c) $d_1=2.7$, (d) $d_1=2.79$, (e) $d_1=3.$, (f) $d_1=3.5$, (g) $d_1=5.$, (h) $d_1=6.$, (i) $d_1=7.$, (j) $d_1=8.7$, (k) $d_1=13.$, (l) $d_1=18.$ The  solid lines are of second-order, the doted ones are of first-order. Filled circles are the tricritical points, empty circles are the $B^3$ multicritical points, filled squarres are the $BA^2$ critical end-points and the empty squarre is the $A^3$ end-point. 
\item[Fig.4] The phase diagrams in the (t,$d_2$) plane for (a) $M=2$ and $N=9$ ($3-1/M=2.5$), (b)$M=9$ and $N=2$ ($3+1/M=3.11$) .The number accompanying each curve is the value of $d_1$. The  solid lines are of second-order, the doted ones are of first-order. Filled circles are the tricritical points.
\end{description}
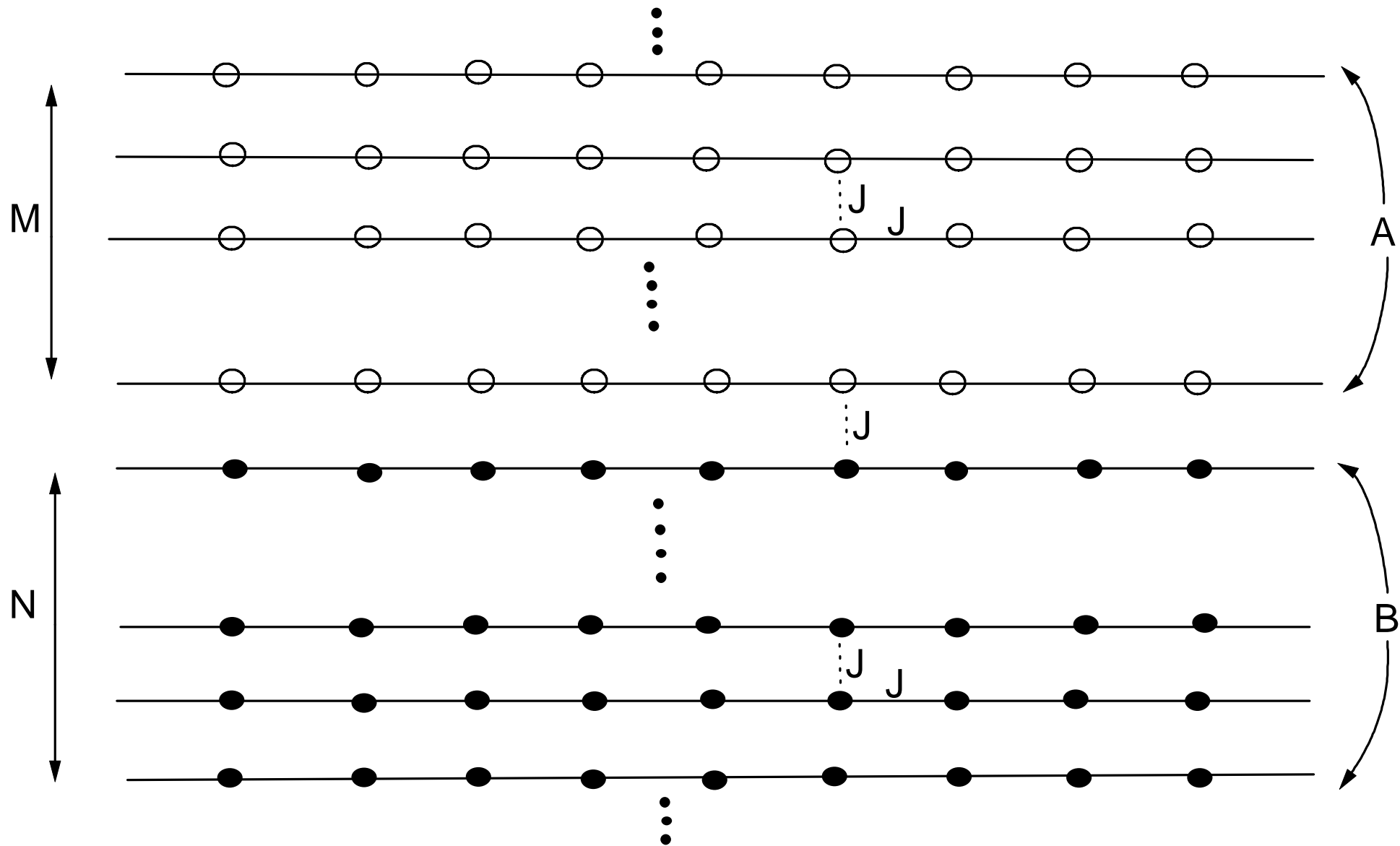
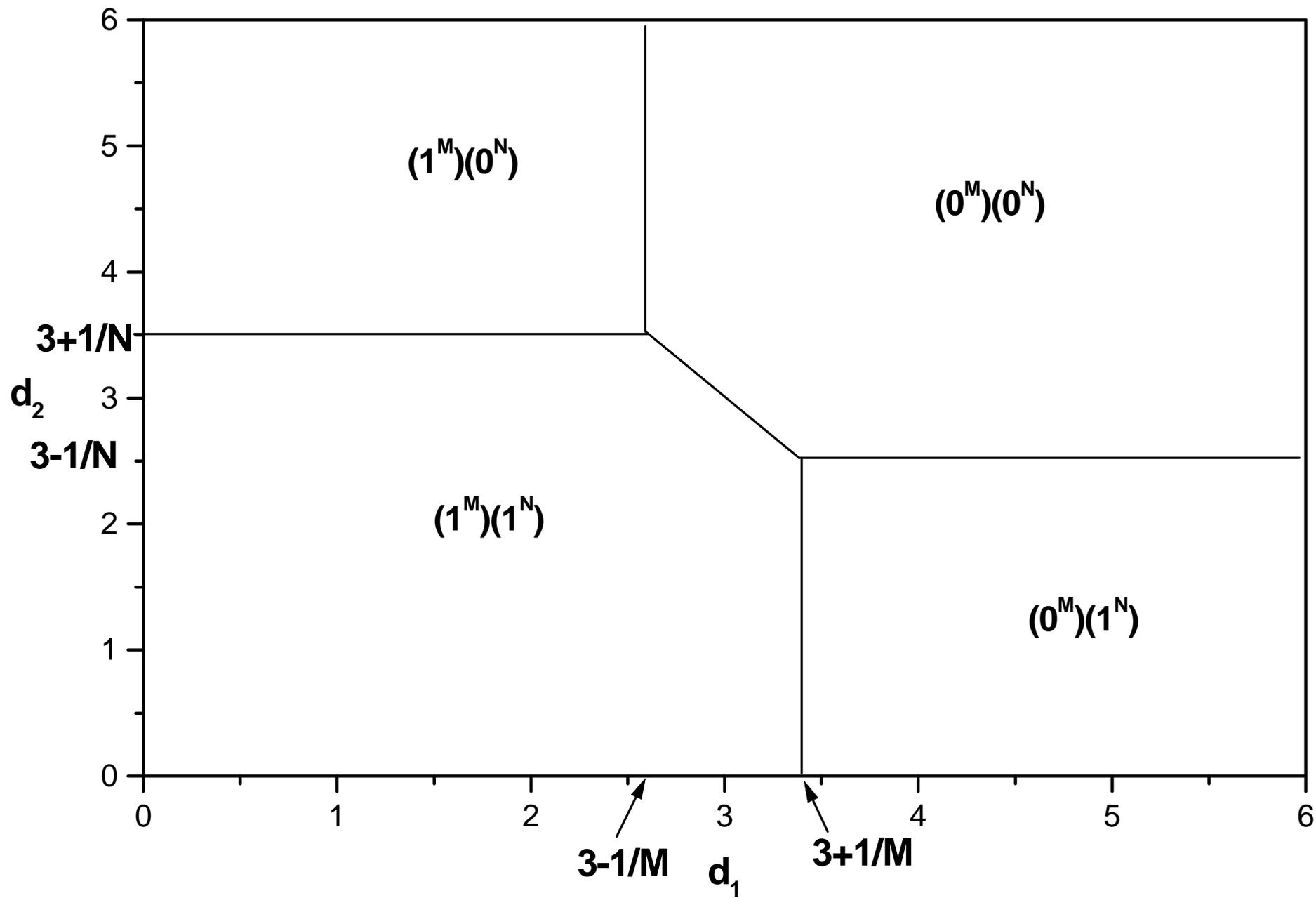
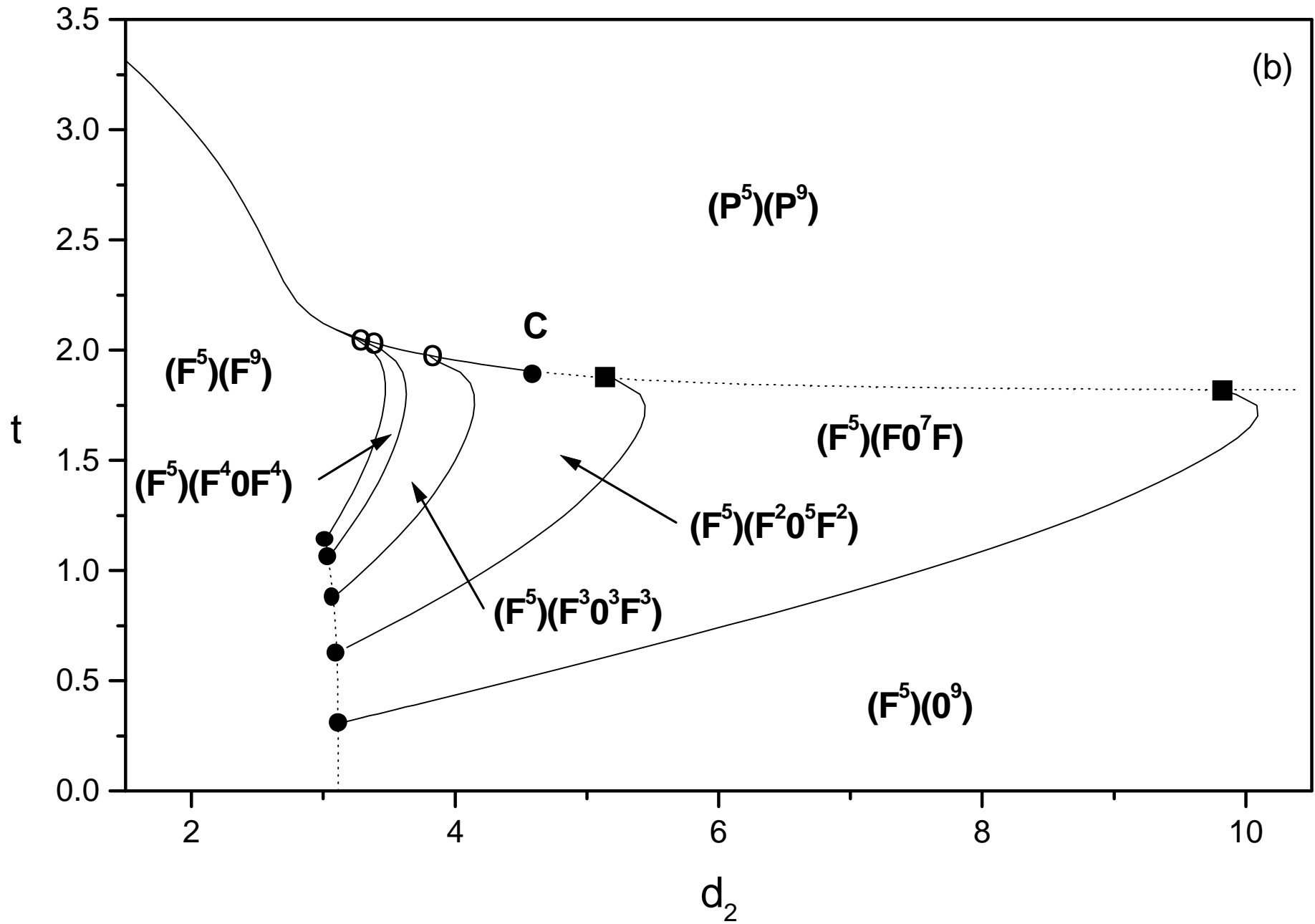
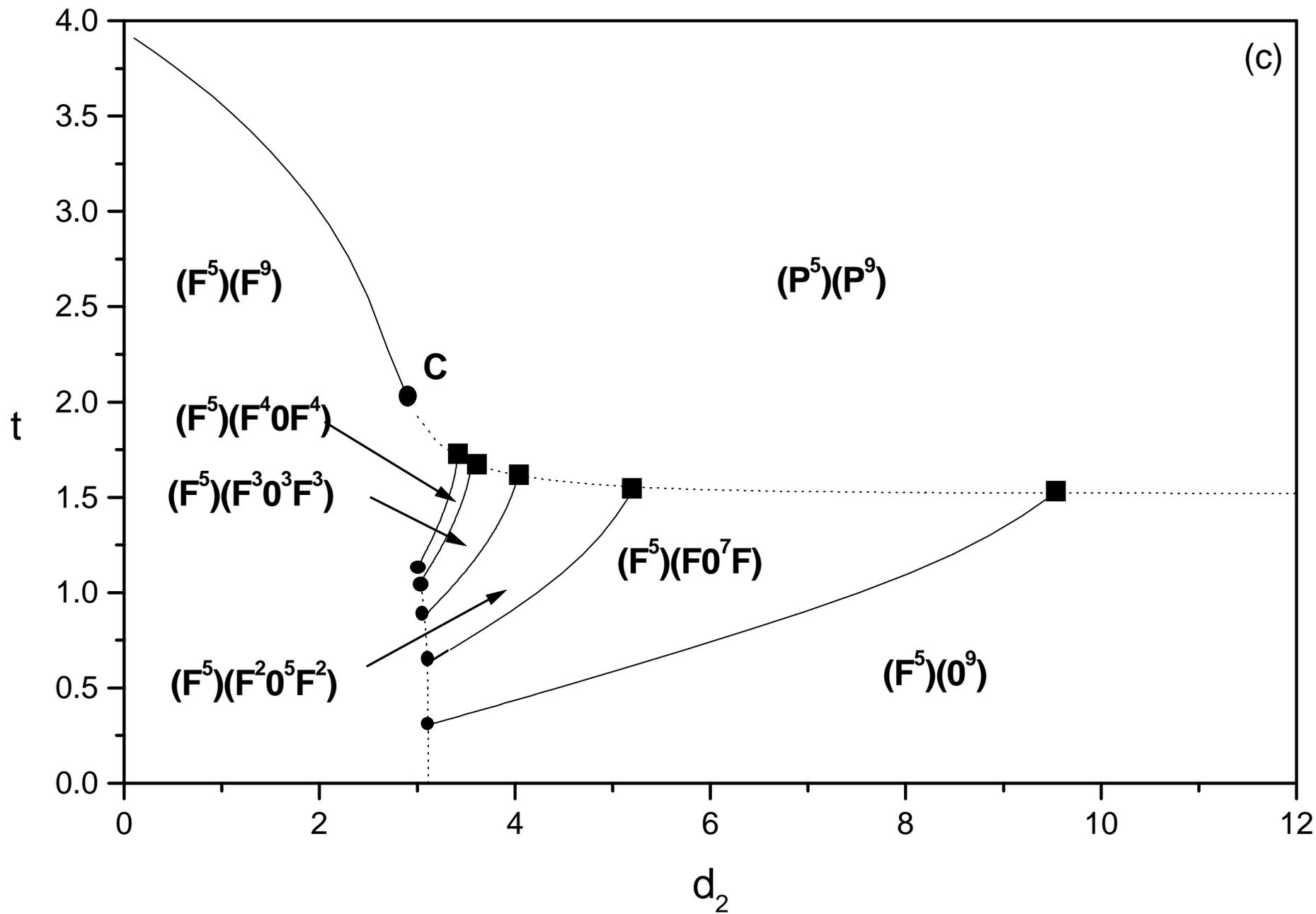
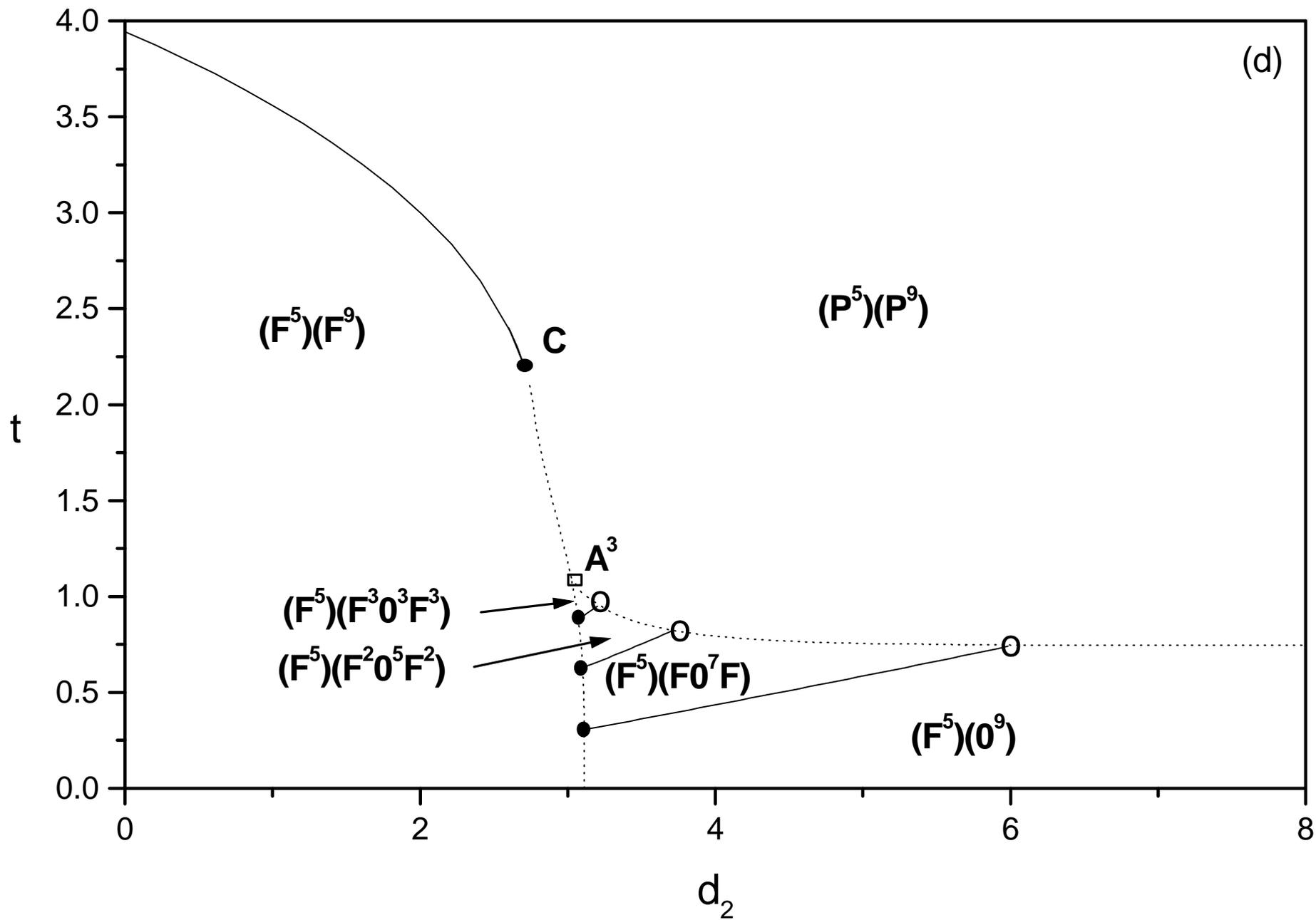
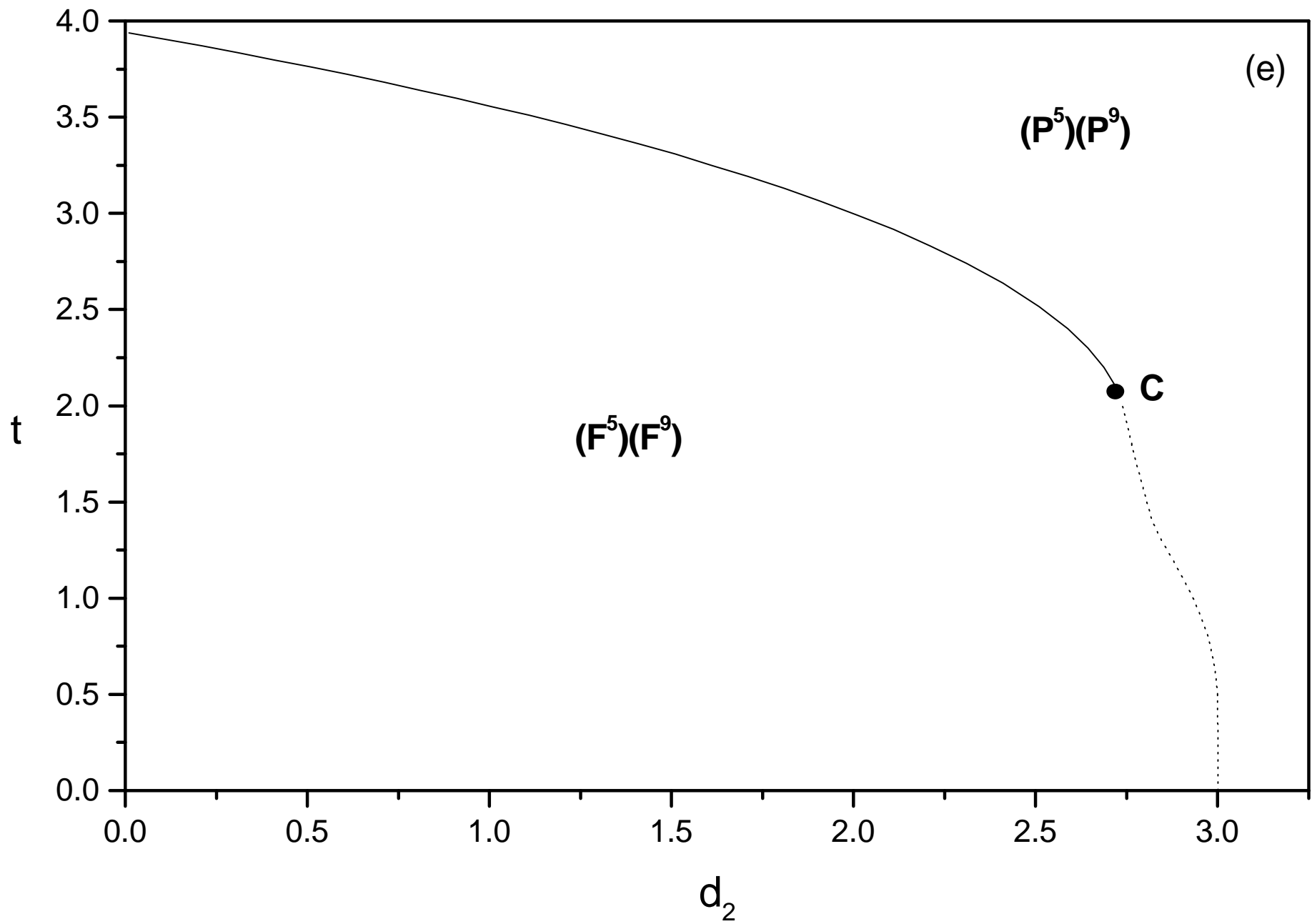
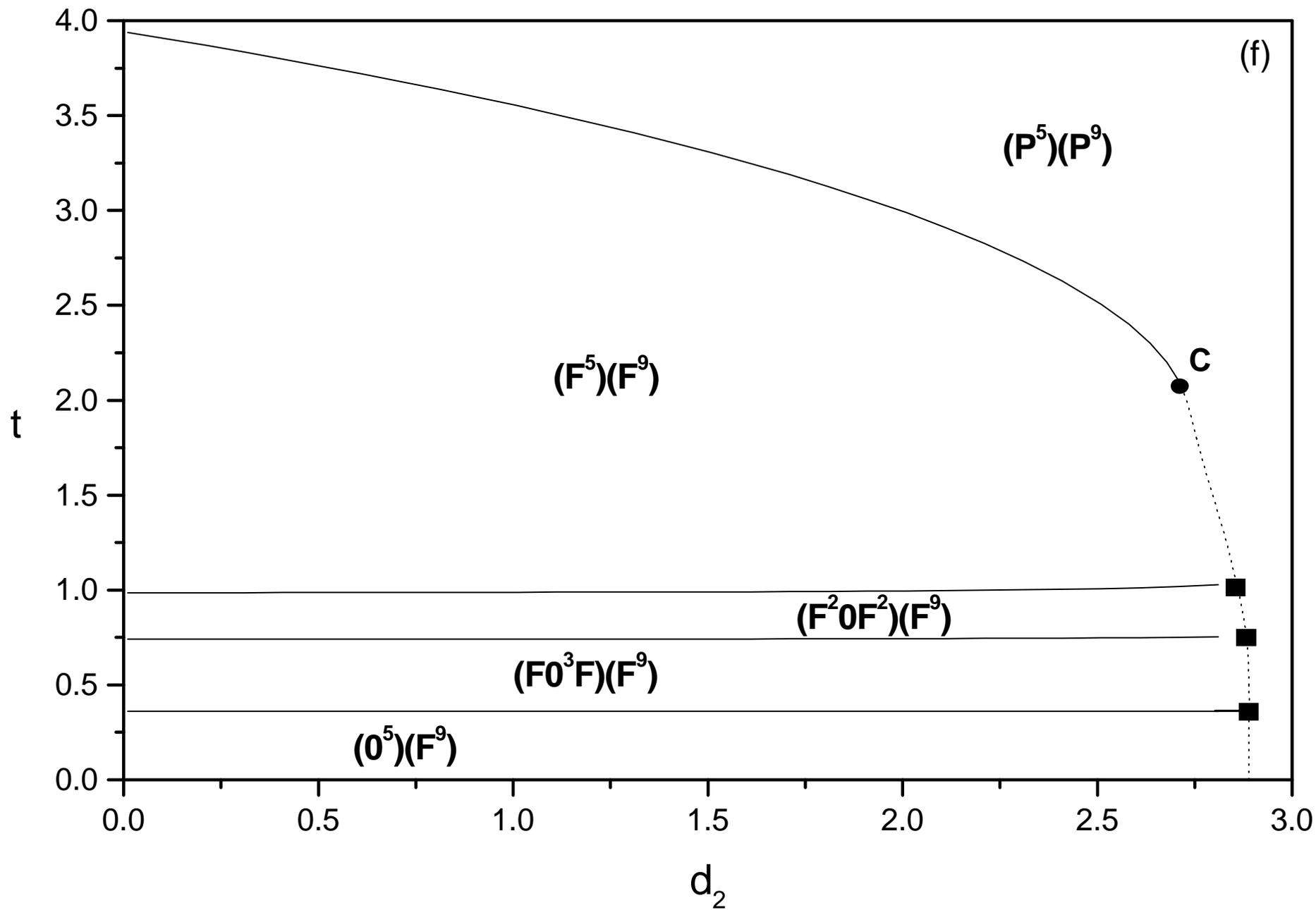
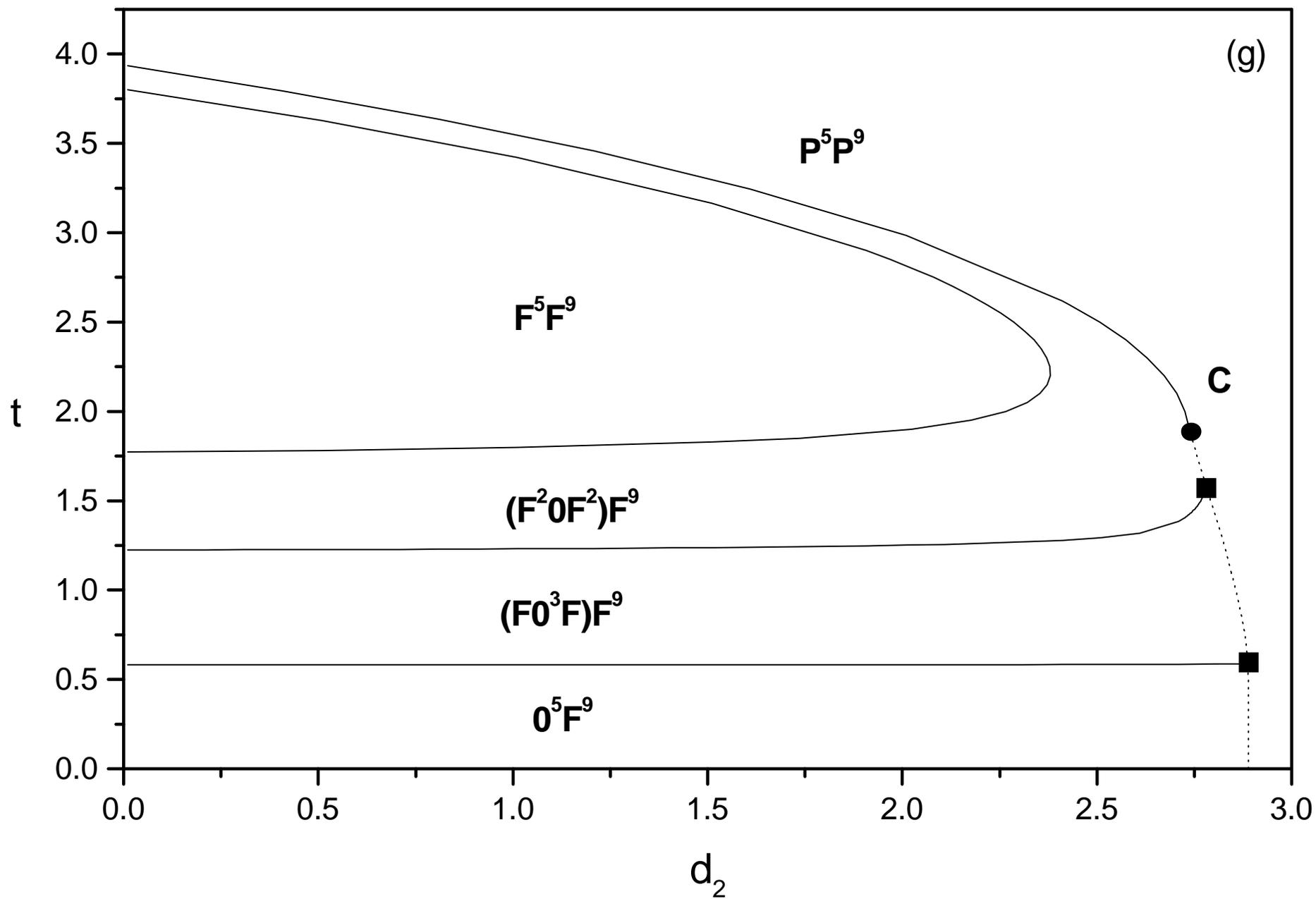
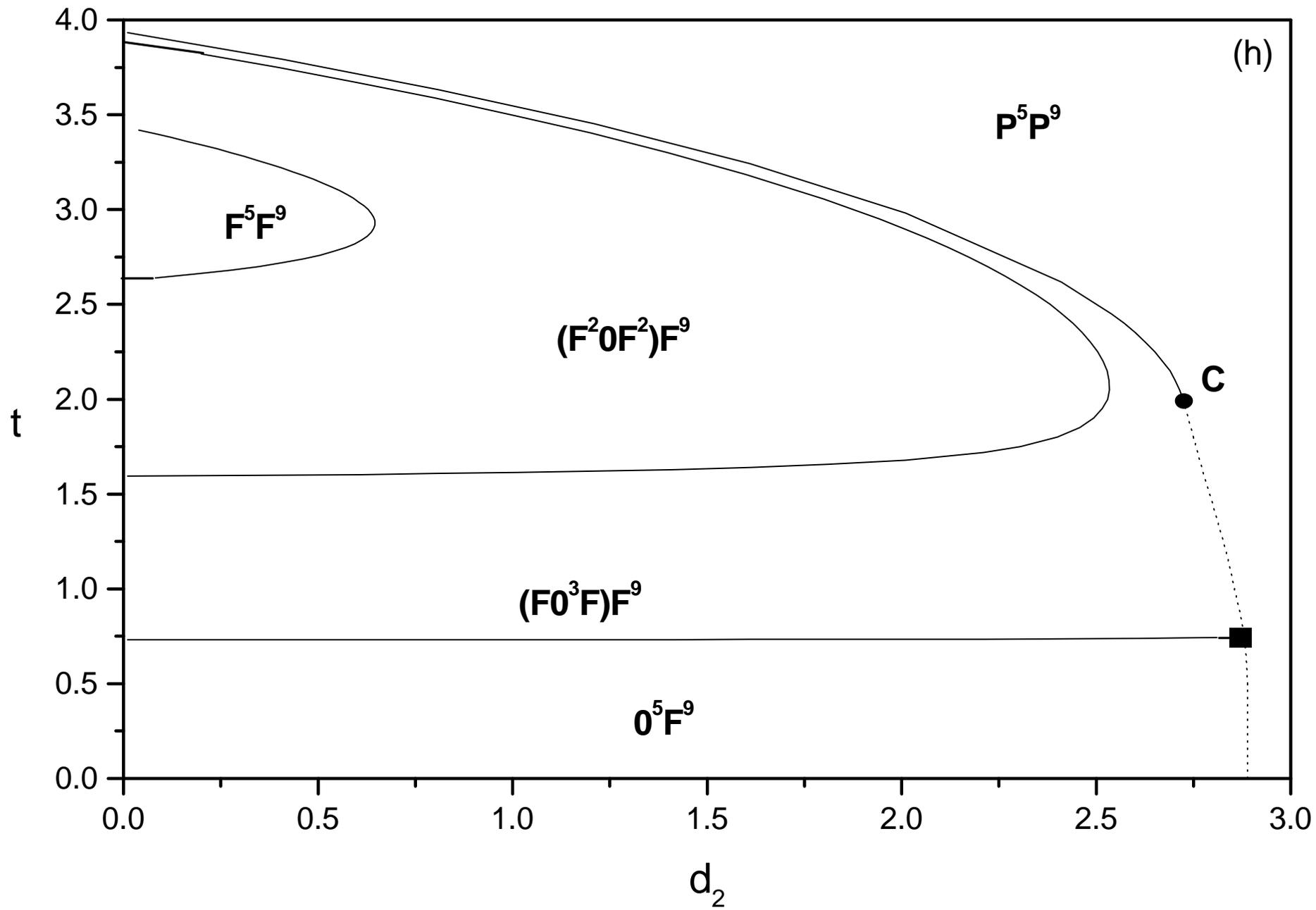
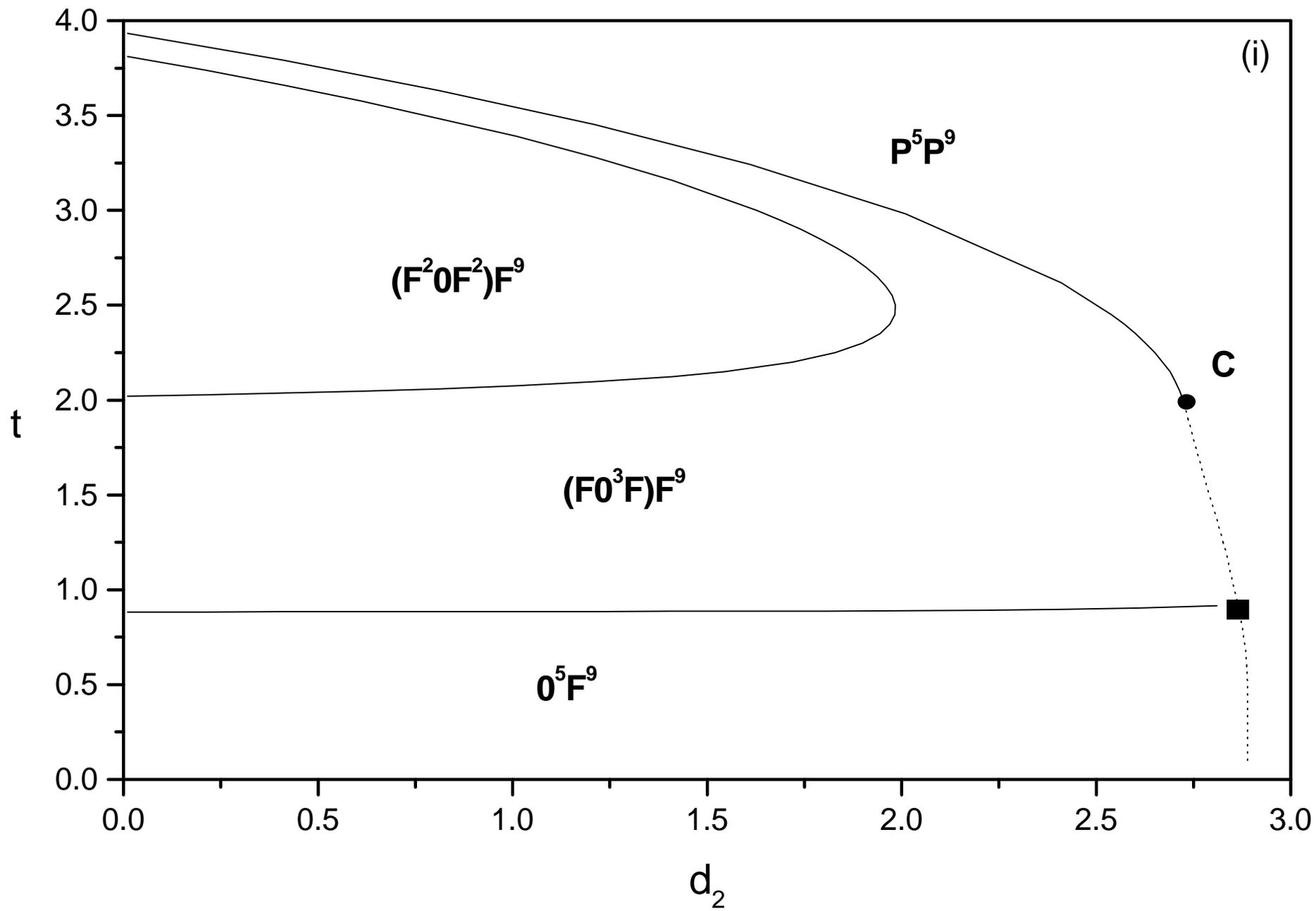
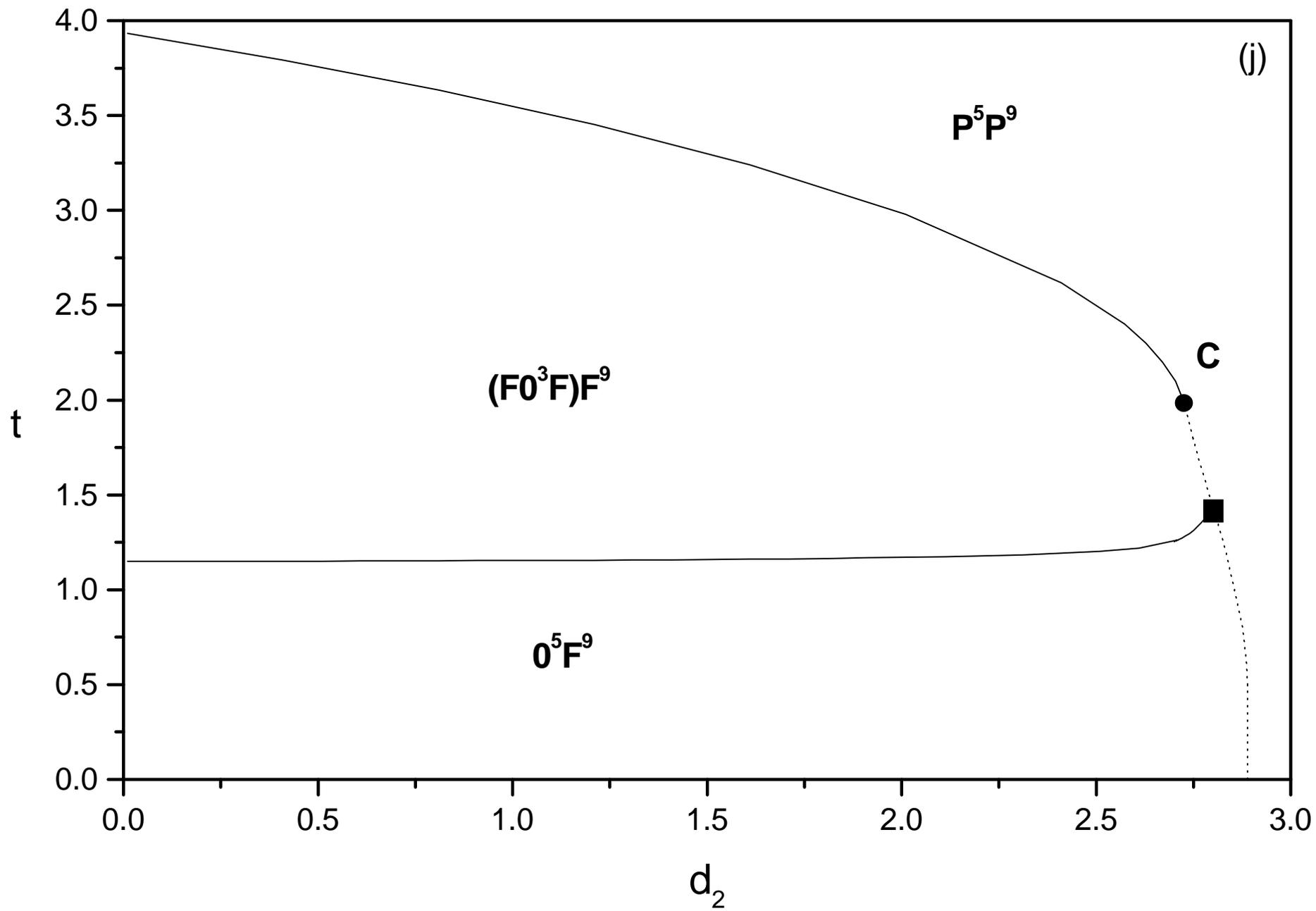
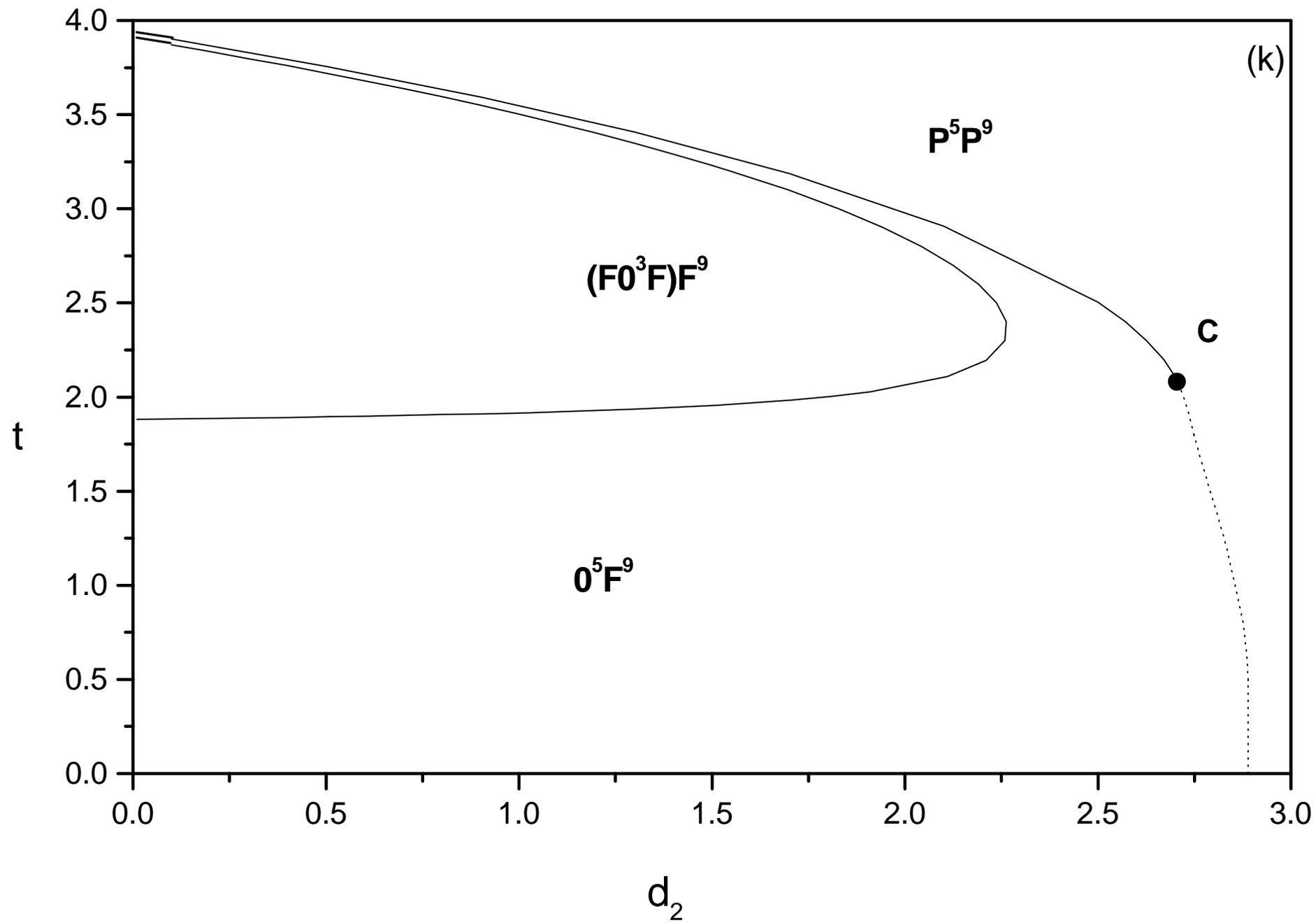
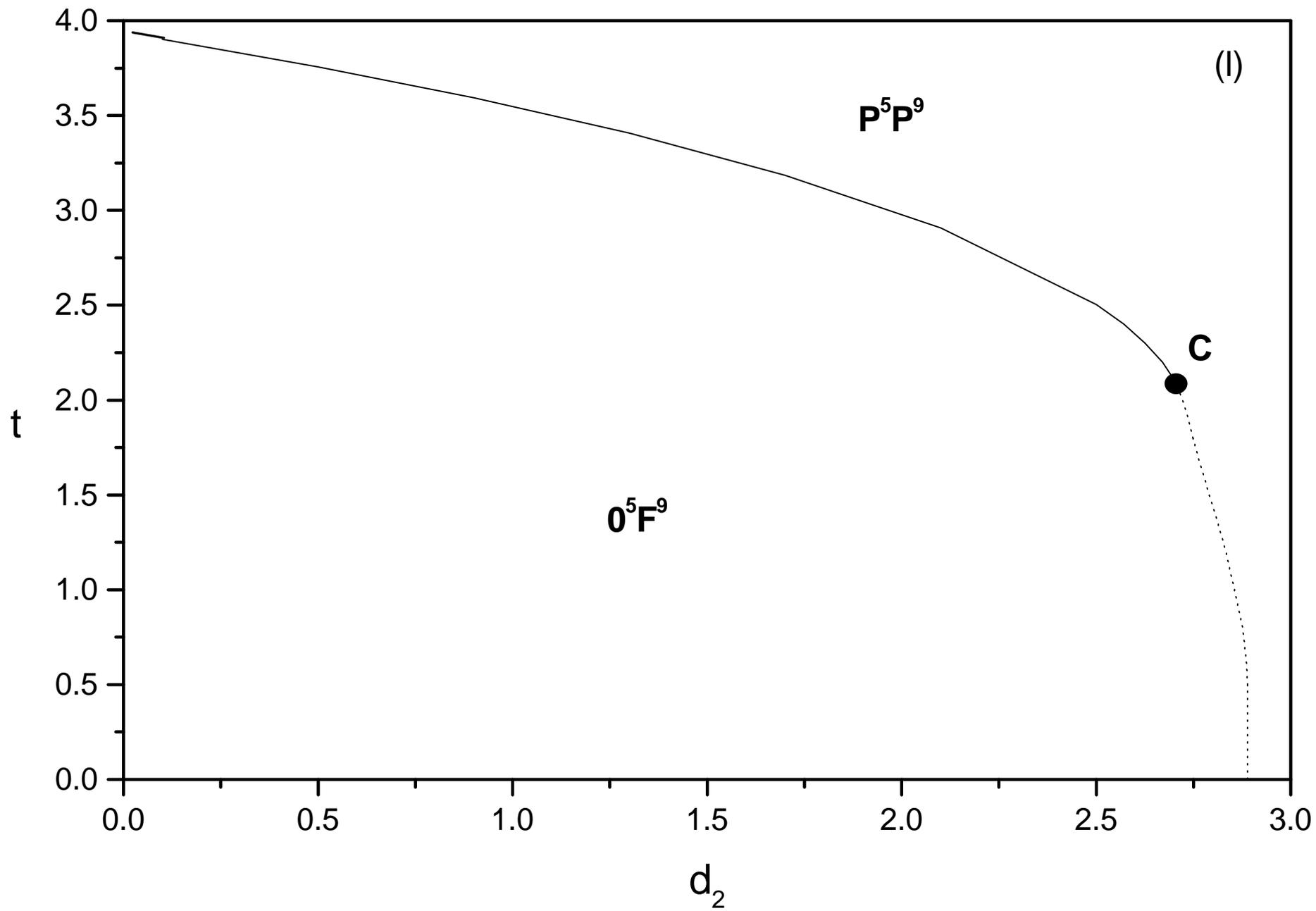
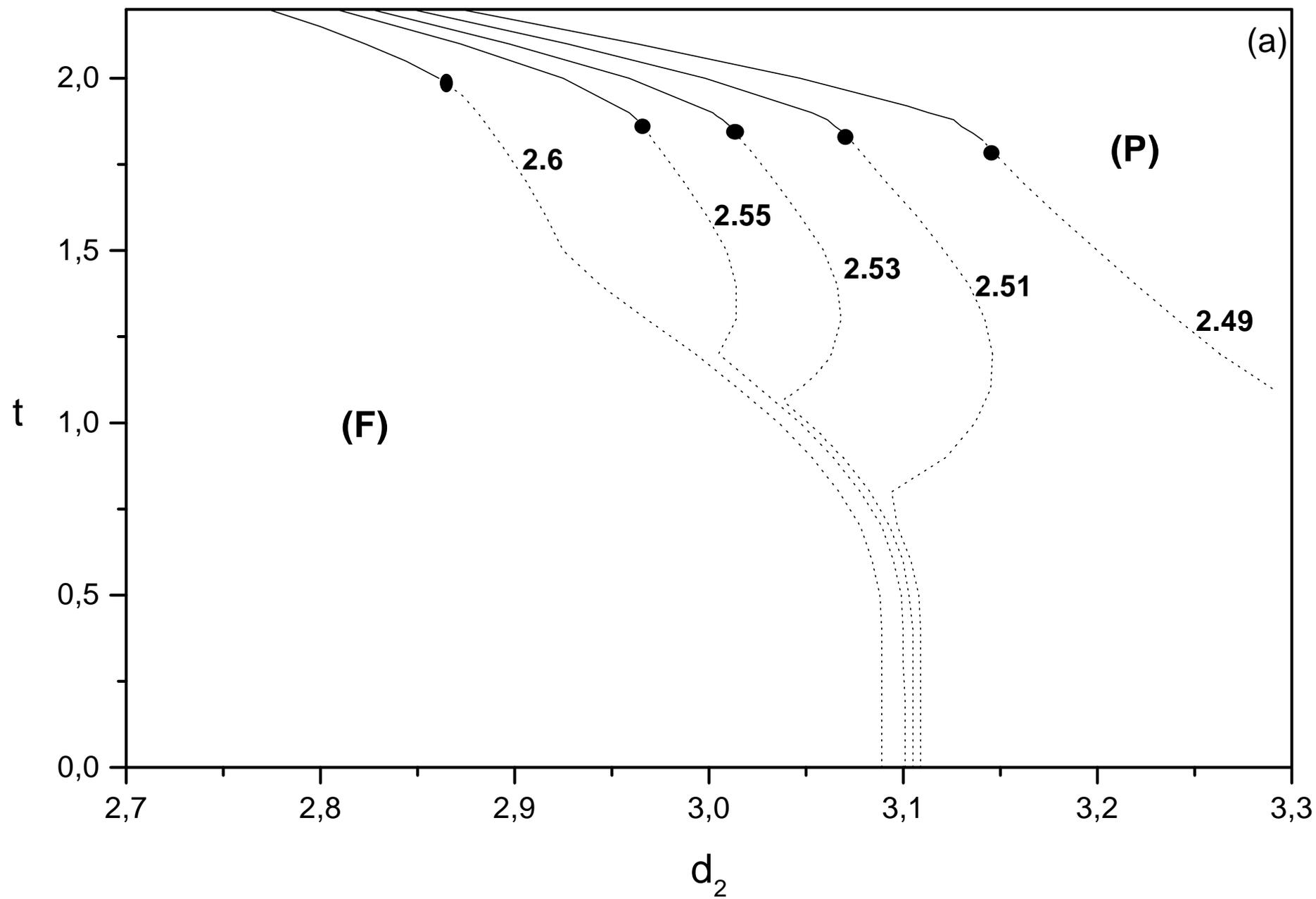
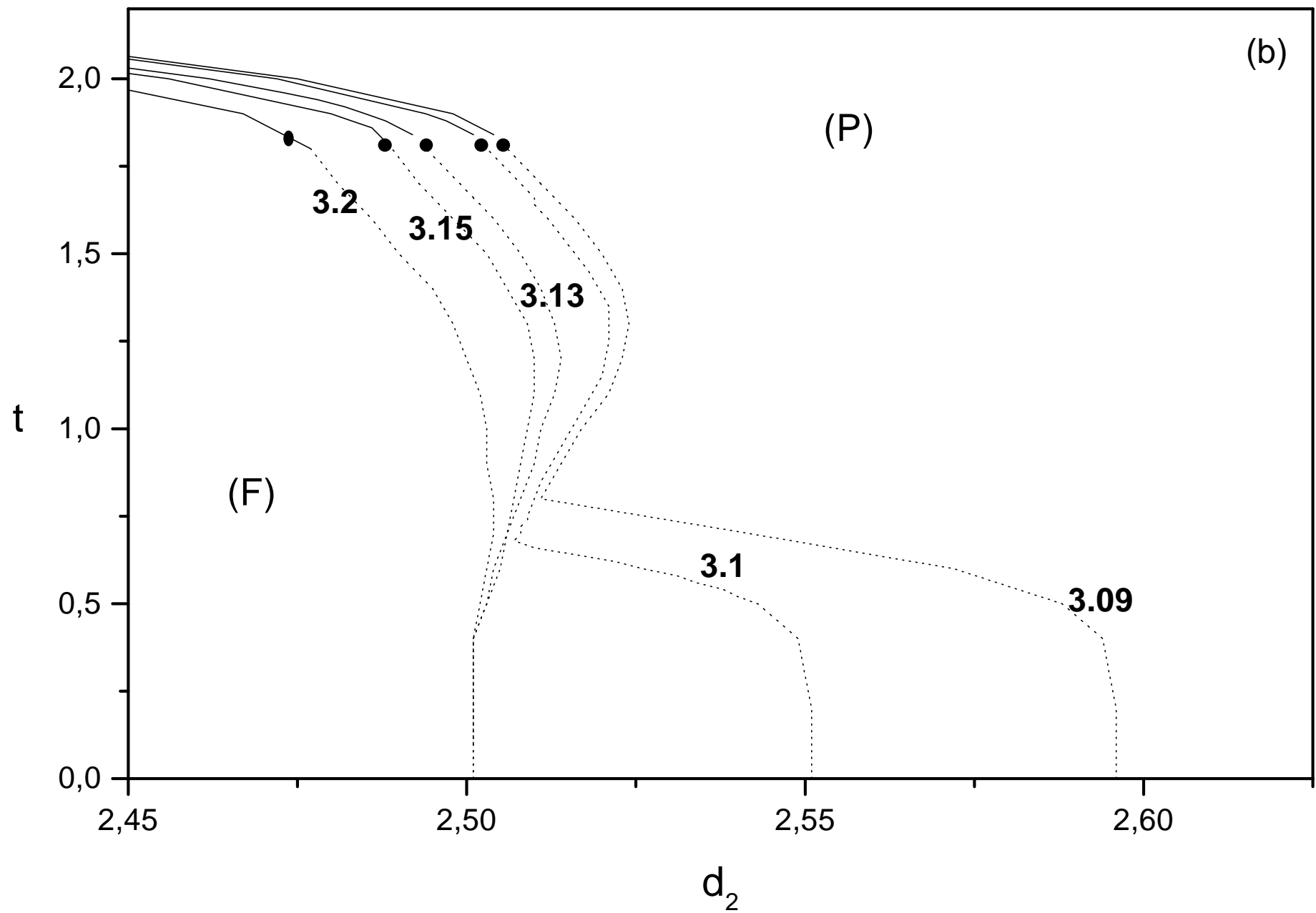
\end{document}